\documentclass[aps,twocolumn,showpacs,superscriptaddress]{revtex4}
\usepackage{epsf}
\usepackage{multirow}

\begin{document}

\title{Search on a Hypercubic Lattice through a Quantum Random Walk: II. d=2}
\author{Apoorva Patel}
\email{adpatel@cts.iisc.ernet.in}
\affiliation{Centre for High Energy Physics,
             Indian Institute of Science, Bangalore-560012, India}
\affiliation{Supercomputer Education and Research Centre,
             Indian Institute of Science, Bangalore-560012, India}
\author{K.S. Raghunathan}
\email{ksraghu@yahoo.com}
\affiliation{Centre for High Energy Physics,
             Indian Institute of Science, Bangalore-560012, India}
\author{Md. Aminoor Rahaman}
\email{aminoorrahaman@yahoo.com}
\affiliation{Supercomputer Education and Research Centre,
             Indian Institute of Science, Bangalore-560012, India}
\date{\today}

\begin{abstract}
\noindent
We investigate the spatial search problem on the two-dimensional square
lattice, using the Dirac evolution operator discretised according to the
staggered lattice fermion formalism. $d=2$ is the critical dimension for
the spatial search problem, where infrared divergence of the evolution
operator leads to logarithmic factors in the scaling behaviour. As a result,
the construction used in our accompanying article \cite{dgt2search}
provides an $O(\sqrt{N}\ln N)$ algorithm, which is not optimal.
The scaling behaviour can be improved to $O(\sqrt{N\ln N})$ by cleverly
controlling the massless Dirac evolution operator by an ancilla qubit,
as proposed by Tulsi \cite{tulsi}. We reinterpret the ancilla control
as introduction of an effective mass at the marked vertex, and optimise
the proportionality constants of the scaling behaviour of the algorithm
by numerically tuning the parameters.
\end{abstract}
\maketitle

\section{Two-dimensional Spatial Search}

The spatial search problem is to find a marked object from an unsorted
database of size $N$ spread over distinct locations, with the restriction
that one can proceed from any location to only its neighbours while
inspecting the objects. Classical algorithms for unsorted database search
are $O(N)$, but quantum algorithms working with a superposition of states
do better. The quantum spatial search problem was first formulated on a
single hypercube in Ref.\cite{hypsrch1}, and its two-dimensional version
has received particular attention due to its critical nature
\cite{gridsrch1,gridsrch2,tulsi,hexsrch}.
In our accompanying article \cite{dgt2search}, we have investigated the
spatial search problem for $d>2$ hypercubic lattices using the massless
Dirac evolution operator, and obtained close to the optimal scaling
behaviour of Grover's algorithm \cite{grover}. We have also developed a
broad picture of how the dimension of the database influences the spatial
search problem, in analogy with statistical mechanics of critical phenomena.
In case of $d\leq2$, the evolution operator arising from the massless
Dirac operator is infrared divergent (as $\int d^d k / k^2$ in the
continuum formulation). That slows down spatial search algorithms by
logarithmic factors in $d=2$ \cite{gridsrch1,gridsrch2,tulsi,hexsrch},
compared to the optimal scaling form, and our algorithm suffers the same
fate. In this article, we modify our algorithm by introducing an effective
mass in the Dirac evolution operator, and demonstrate how that overcomes
the infrared hurdle and improves the scaling behaviour.

Discrete time algorithms for the spatial search problem are based on
random walks. In quantum random walks \cite{qrw}, the state amplitude
distribution is unitarily evolved, such that amplitude at each vertex
gets redistributed over itself and its neighbours at every time step.
Quantum superposition allows simultaneous exploration of multiple
possibilities, and this technique has become part of a variety of
algorithms (Ref.\cite{qrwrev} provides an introductory overview).

We study the specific case of search for a marked vertex on a square lattice
with $N=L^2$ vertices. In our algorithmic strategy \cite{grover_strategy},
the free Dirac Hamiltonian,
\begin{equation}
H_{\rm free} = -i\vec{\alpha}\cdot\vec{\nabla} + \beta m ~,
\label{diracham}
\end{equation}
diffuses the amplitude distribution around the lattice, while the potential
attracting the amplitude distribution toward the marked vertex provides the
binary oracle,
\begin{eqnarray}
&& V = V_0 \delta_{\vec{x},0} ~,~~ e^{-iV_0 \tau} = -1 \cr
&\Longrightarrow& R = e^{-iV\tau} = I - 2 |\vec{0}\rangle\langle\vec{0}| ~.
\end{eqnarray}
Upon discretising the Hamiltonian according to the staggered lattice fermion
formalism \cite{staggered}, the anticommuting Dirac matrices are reduced to
location dependent signs,
\begin{equation}
\alpha_n \longrightarrow \prod_{j=1}^{n-1} (-1)^{x_j} ~,~~
\beta \longrightarrow \prod_{j=1}^{2} (-1)^{x_j} \beta ~.
\label{pmsigns}
\end{equation}

To construct discrete and local time evolution, we need to exponentiate
the terms in the Hamiltonian such that the resultant unitary operators
are local. The mass and potential terms are single vertex local terms,
and so can be exponentiated easily. To obtain local exponential of the
kinetic term connecting neighbouring vertices, we separate the ``odd''
and ``even'' parts of the Hamiltonian on the bipartite square lattice,
and exponentiate each block-diagonal Hermitian part separately \cite{qwalk2}:
\begin{equation}
H_{\rm free} = H_o + H_e + H_m ~,~~ U = \exp(-iH\tau) ~.
\end{equation}
The $4\times4$ blocks of the Hamiltonian, corresponding to hypercubes with
coordinate labels $\{0,1\}^{\otimes 2}$, are (in terms of tensor products
of Pauli matrices)
\begin{equation}
H_o^B = -{\textstyle{1\over2}}
        ( I\otimes\sigma_2 + \sigma_2\otimes\sigma_3 ) ~,
\label{hblock}
\end{equation}
and $H_e^B=-H_o^B$ when operating on a hypercube with all coordinates
flipped in sign. The factors appearing in the Trotter's formula for the
discrete time evolution operator then become
\begin{equation}
U_{o(e)} = cI - is\sqrt{2}H_{o(e)} ~,~~ 
\end{equation}
with $s=\sin(\tau/\sqrt{2})$ and $|c|^2 + |s|^2 = 1$.
We choose the unitary quantum walk operator to be
\begin{equation}
W = U_e U_o = \exp(-i(H_o+H_e)\tau) + O(\tau^2) ~.
\end{equation}

Our initial state for the spatial search problem is the unbiased uniform
superposition state,
$|s\rangle=\sum_x |\vec{x}\rangle/\sqrt{N}$.
For $m=0$, our algorithm alternates between the oracle and the walk
operators, yielding the evolution
\begin{equation}
\psi(\vec{x};t_1,t_2) = [W^{t_1} R]^{t_2} \psi(\vec{x};0,0) ~.
\label{evolsearch}
\end{equation}
Here $t_2$ is the number of oracle calls, and $t_1$ is the number of walk
steps between the oracle calls. Both have to be optimised, depending on the
size of the lattice, to find the quickest solution to the search problem.

With iterative unitary operations, spatial search algorithms produce results
periodic in time. Unlike Grover's algorithm, however, the maximum probability
of being at the marked vertex, $P$, does not reach the value $1$. Augmenting
the algorithm by the amplitude amplification procedure \cite{brassard}, the
marked vertex can be found with probability $\Theta(1)$, and the overall
complexity of the algorithm is then characterised by the effective number of
oracle calls $t_2/\sqrt{P}$.

\section{Improving the Algorithm}

We have argued \cite{dgt2search} that spatial search in $d$ dimensions
obeys two lower bounds,
\begin{equation}
t_2 ~\geq~ {\rm max}\{d N^{1/d}, \pi\sqrt{N}/4\} ~,
\label{bounds}
\end{equation}
following from distinct physical principles of special relativity and
unitarity, respectively. The bounds cross in the critical dimension $d=2$,
where logarithmic corrections to scaling behaviour are expected in analogy
with critical phenomena in statistical mechanics. Such logarithmic slow down
factors have been observed in earlier works \cite{gridsrch1,gridsrch2,tulsi},
and we want to suppress them as much as possible by suitably adjusting the
evolution Hamiltonian.

When $m=0$, the quantum random walk provides the fastest diffusion,
minimising $t_2$. But $m=0$ also makes the evolution operator infrared
divergent in two dimensions, which spreads out the amplitude distribution
in the Hilbert space too much and decreases $P$. Introduction of $m\ne0$
would slow down the diffusion, but would also regulate the infrared
divergence through $k^2 \rightarrow k^2 + m^2$. For small enough $m$,
the diffusion speed (and hence $t_2$) may not change much, but substantial
modification of the contribution of $|\vec{k}| \lesssim m$ modes can alter
the behaviour of $P$. In such a case, an optimal value of $m$ can be
obtained by trading off the increase in $t_2$ against the increase in $P$.
For a finite lattice, the lattice size acts as the infrared cutoff, and so
the optimal value of $m$ is expected to be a function of the database size $N$.

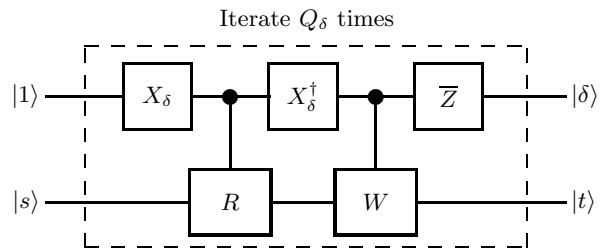
\begin{figure}[t]
\setlength{\unitlength}{1pt}
\begin{picture}(250,100)(10,0)
\thicklines
\put(33,25){\makebox(0,0)[r]{$|s\rangle$}}
\put(33,65){\makebox(0,0)[r]{$|1\rangle$}}
\put(234,25){\makebox(0,0)[l]{$|t\rangle$}}
\put(234,65){\makebox(0,0)[l]{$|\delta\rangle$}}
\put(35,65){\line(1,0){30}}
\put(90,65){\line(1,0){30}}
\put(145,65){\line(1,0){30}}
\put(200,65){\line(1,0){32}}
\put(65,53){\framebox(25,24){$X_{\delta}$}}
\put(105,65){\circle*{6}}
\put(120,53){\framebox(25,24){$X_{\delta}^{\dagger}$}}
\put(160,65){\circle*{6}}
\put(175,53){\framebox(25,24){$\overline{Z}$}}
\put(35,25){\line(1,0){55}}
\put(120,25){\line(1,0){25}}
\put(175,25){\line(1,0){57}}
\put(90,13){\framebox(30,24){$R$}}
\put(105,62){\line(0,-1){25}}
\put(145,13){\framebox(30,24){$W$}}
\put(160,62){\line(0,-1){25}}
\thinlines
\put(50,8){\dashbox{5}(167,76){}}
\put(134,90){\makebox(0,0)[b]{{Iterate $Q_{\delta}$ times}}}
\end{picture}
\caption{Logic circuit diagram for Tulsi's controlled quantum spatial
search algorithm. $R$ and $W$ are the binary oracle and the massless
Dirac walk operator, respectively. We use the generalisation with
$W \longrightarrow W^{t_1}$.}
\end{figure}

\subsection{Tulsi's Algorithm}

Tulsi constructed an algorithm possessing the previously described properties,
by controlling the evolution operators using an ancilla qubit \cite{tulsi}.
His scheme is illustrated in Fig.1, where the ancilla operators are
\begin{equation}
X_\delta = \pmatrix{ \cos\delta & \sin\delta \cr
                    -\sin\delta & \cos\delta } ~,~~
\overline{Z} = \pmatrix{ -1 & 0 \cr
                          0 & 1 } ~,
\end{equation}
and the algorithm evolves the initial state $|1\rangle\otimes|s\rangle$
to the target state $|\delta\rangle\otimes|t\rangle$ with
$|\delta\rangle = X_\delta^\dag|1\rangle$.

For $\delta=0$, Tulsi's algorithm reduces to spatial search using the
massless Dirac walk operator, which finds the marked vertex with
probability $P_0=\Theta(1/\ln N)$ using $Q_0=\Theta(\sqrt{N\ln N})$
oracle calls \cite{gridsrch1,gridsrch2}. Tulsi showed that with
$\cos\delta = \Theta(1/\sqrt{\ln N})$, the algorithm increases the
probability of finding the marked vertex to $P_\delta=\Theta(1)$ without
changing the scaling of oracle calls $Q_\delta=\Theta(\sqrt{N\ln N})$
\cite{tulsi}. More explicitly, the algorithm largely confines the evolution
of the quantum state to a two-dimensional subspace of the $N$-dimensional
Hilbert space, whereby
\begin{equation}
P_\delta = {1 \over 2^d B_\delta^2} ~,~~
Q_\delta = {\pi B_\delta \sqrt{N/2^d} \over 4\cos\delta} ~,
\label{PQvalues}
\end{equation}
\begin{equation}
B_\delta^2 = 1+(B^2-1)\cos^2\delta ~.
\label{Bdelta}
\end{equation}
Here $B \equiv B_0 = \Theta(\sqrt{\ln N})$ is a second moment constructed
from the eigenspectrum of $W$, and characterises the infrared divergence
of the problem. We have also included explicit factors of $2^d$ that are
appropriate for spatial search with staggered fermions, where different
vertices of an elementary hypercube correspond to different degrees of
freedom, and essentially only the degree of freedom corresponding to the
marked vertex participates in the search process \cite{dgt2search}.

The optimal value of the ancilla control parameter is obtained by
minimising the algorithmic complexity (where factors of $2^d$ cancel):
\begin{equation}
(\cos\delta)_{\rm opt} = (B^2-1)^{-1/2} \approx 1/B ~,
\label{optdelta}
\end{equation}
\begin{equation}
(B_\delta^2)_{\rm opt} = 2 ~,~~
\Big({Q_\delta \over \sqrt{P_\delta}}\Big)_{\rm min}
= {\pi\sqrt{N} B \over 2} ~.
\label{optcomplexity}
\end{equation}

\subsection{Rephrasing of Tulsi's Algorithm}

Tulsi's algorithm is not presented in the Dirac operator language---rather
the ancilla qubit is introduced in an ad hoc manner. Here we relate its
ingredients to the well-known properties of the Dirac operator, in order
to gain some physical insight in to its dynamical behaviour.

Consider two species of Dirac particles: one with only the mass term in
the Hamiltonian (completely at rest), and the other with only the kinetic
term in the Hamiltonian (fully relativistic). Associating the species index
with the ancilla value, we have
\begin{eqnarray}
H_{\rm free}^{(0)} &=& |0\rangle\langle0| \otimes H_m ~, \cr
H_{\rm free}^{(1)} &=& |1\rangle\langle1| \otimes (H_o+H_e) ~.
\end{eqnarray}
Now, pick $m$ to give the farthest unitary evolution, i.e.
\begin{equation}
e^{-i\beta m\tau} = -1 ~.
\end{equation}
(The sign provided by $\beta$ does not matter in this case.)
With these choices, the two species evolve independently with maximum
evolution contrast. The total Hamiltonian then yields the second half of
the iteration in Tulsi's algorithm---schematically,
\begin{eqnarray}
e^{-i\big(H_{\rm free}^{(0)}+H_{\rm free}^{(1)}\big)\tau}
  &\longrightarrow& \pmatrix{-1 & 0 \cr
                            0 & W } \cr
  &=& (c_1 W)\overline{Z} ~=~ \overline{Z}(c_1 W) ~.
\end{eqnarray}

The first half of the iteration in Tulsi's algorithm is a controlled
oracle, conditioned on the state $|\delta\rangle$, i.e.
\begin{equation}
(c_\delta R) = X_\delta^\dagger (c_1 R) X_\delta ~.
\end{equation}
For the vertices $\vec{x}\ne0$ without the potential, it is the identity
operation. That lets the ``$|0\rangle$'' species remain at rest while
the ``$|1\rangle$'' species diffuses at full speed, and there is no
mixing between the two. A consequence is that the amplitudes
$|0\rangle\otimes|\vec{x}\ne0\rangle$ do not change from their
initial value zero. They neither mix with the amplitudes
$|1\rangle\otimes|\vec{x}\ne0\rangle$ nor get any contribution from
the amplitude $|0\rangle\otimes|\vec{x}=0\rangle$ by a walk step.

The potential at $\vec{x}=0$ couples the two species, as per
\begin{equation}
(c_\delta R)_{\vec{x}=0} = Z X_{2\delta} = \overline{Z} X_{2\delta-\pi} ~.
\label{coupling}
\end{equation}
As a result, when the diffusing ``$|1\rangle$'' species reaches the
marked vertex, part of it gets converted to the ``$|0\rangle$'' species
and stops diffusing. At the next iteration, part of the ``$|0\rangle$''
species gets reconverted to the ``$|1\rangle$'' species and starts
diffusing again. The net effect is that the conversions reduce the number
of $W$ operations for walks when they pass through the marked vertex.
The state $|0\rangle\otimes|\vec{x}=0\rangle$ thus acts like a trap, and
the concentration of the quantum state amplitude at the marked vertex 
increases $P_\delta$. Since the conversion between species pauses the walk,
it can be interpreted as an effective mass, but this effective mass is
unusual in the sense that it appears only at the marked vertex.

The determination of the optimal value of $\delta$ requires an analysis
of the eigenspectrum of the operator $W$. As shown by Tulsi \cite{tulsi},
when the evolution is largely confined to a two-dimensional subspace of
the Hilbert space, $(\cos\delta)_{\rm opt}=\Theta(1/\sqrt{\ln N})$ and
$\delta\approx\pi/2$. Note that for $\delta=\pi/2$, Eq.(\ref{coupling})
implies that the two species decouple completely. So the mixing of species
per iteration is tiny. Also, the target state is essentially the trap state,
$|\delta\rangle\otimes|\vec{x}=0\rangle \approx |0\rangle\otimes|\vec{x}=0\rangle$,
which is reached from the initial state $|1\rangle\otimes|s\rangle$ with
$\Theta(1)$ probability by accumulating the mixing of species over many
iterations. The trapping of the quantum amplitude, resulting from an
effective mass at a specific location, is a noteworthy feature---it can
be an important ingredient in physical applications of the algorithm.

\section{Simulation Results}

We carried out numerical simulations of our quantum spatial search algorithm,
with a single marked vertex, and both with and without ancilla control. The
choice to keep the walk matrices ($U_o$ and $U_e$), as well as the ancilla
operators ($X_\delta$ and $\overline{Z}$), real was convenient for numerical
simulations.

As in the case of our $d>2$ simulations \cite{dgt2search}, we first scanned
for the best values of the parameters $s$ and $t_1$ to optimise the algorithm.
We again found that correlated $s-t_1$ pairs simultaneously maximise $P$
and minimise the corresponding value of $t_2$. With increasing $t_1$,
$P$ increases somewhat and $t_2$ decreases slightly, but they are minor
improvements. The major difference was observed in the dependence of
the optimal parameters on the lattice size. Without ancilla control and for
fixed $t_1$, the optimal $P$ decreases and the optimal $s$ increases with
increasing $L$. The variable $\theta = \sqrt{2}t_1\sin^{-1}s = t_1\tau$ is
somewhat larger then $\pi$ (the value found in case of $d>2$ \cite{dgt2search}),
and increases with increasing $L$. This dependence on the lattice size is
a consequence of the infrared divergence, whereby all the spatial modes do
not contribute to the search process with equal strength. On the other hand,
with ancilla control and for fixed $t_1$, the optimal values of $P$ and $s$
show little dependence on $L$. The variable $\theta$ is still a bit larger
than $\pi$, but it is more or less a constant. These features indicate that
the infrared divergence of the spatial modes is brought under control.

For concreteness in further analysis, we stuck to the parameter choice
$s=1/\sqrt{2}$ and $t_1=3$, which is close to the optimal choice, both
with and without ancilla control. It is not easy to theoretically estimate
$B$, and hence determine the ideal proportionality constant between the
control parameter $\cos\delta$ and $1/\sqrt{\ln N}$. To figure that out
numerically, we performed simulations with three different choices:
$\cos\delta=\sqrt{1/\ln N}, \sqrt{4/\ln N}, \sqrt{8/\ln N}$.

\begin{table}[t]
\caption{Fit parameters for peak probability without ancilla}
\vspace{2mm}
\begin{tabular}{|c|c|c|c|c|c|}\hline
 $s$ & $t_1$ & $L$ & $a_1$ & $b_1$ & Error \\ \hline
$1\over\sqrt{2}$ & 3 & 512---16384 & 2.607 & -12.76 & 2.42$\times$$10^{-3}$ \\ \hline
\end{tabular}
\end{table}

\begin{figure}[b]
\epsfxsize=9cm
\centerline{\epsfbox{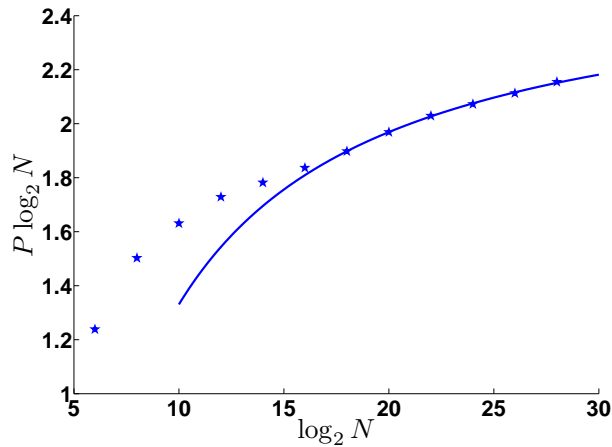}}
\caption{(Color online) Peak probability at the marked vertex as a function
of database size without ancilla control. We used $t_1=3$, and the curve is
the fit $P\log_2 N = a_1 + (b_1/\log_2 N)$.}
\end{figure}

Our results for the dependence of the peak probability on the lattice size
are displayed in Figs.2 and 3. As anticipated, $P$ decreases logarithmically
as $1/\ln N$ without ancilla control, and asymptotically approaches a
constant with ancilla control. Even the behaviour of the subleading
correction changes from $1/\ln L$ without ancilla control to $1/L$ with
ancilla control, reconfirming that ancilla control indeed eliminates
infrared divergence of $P$. The values of our fit parameters are presented
in Tables I and II, where error refers to the root-mean-square (rms)
deviation of the data from the fit.  In Table II, we have also included
the related values of $B_\delta$ following from Eq.(\ref{PQvalues}).

Next we display our results for the dependence of the number of oracle calls
on the lattice size in Fig.4. As expected, $t_2$ increases with decreasing
$\cos\delta$, as species conversions slow down the walk. But this slow down
is only by a multiplicative factor, and $t_2/\sqrt{N\ln N}$ asymptotically
approaches a constant in all cases. Furthermore, the subleading correction
is well parametrised as $1/L$, suggesting that $t_2$ is less affected by
infrared divergence of the problem than $P$ is. There is some oscillatory
pattern in the data \cite{oscillatory}, while the approach to the asymptotic
behaviour becomes smoother with decreasing $\cos\delta$. The values of our
fit parameters are presented in Table III. Again error refers to the
rms deviation of the data from the fit, and the related values of
$B_{\delta\ne0}$ following from Eq.(\ref{PQvalues}) are included.

\begin{table}[t]
\caption{Fit parameters for peak probability with ancilla}
\vspace{2mm}
\begin{tabular}{|c|c|c|c|c|c|c|c|}\hline
 $s$ & $t_1$ & $L$ & $\cos \delta$ & $a_1$ & $b_1$ & Error & $B_\delta$ \\ \hline
\multirow{3} {*}{$1\over\sqrt{2}$} & \multirow{3} {*}3 & \multirow{3} {*}{256---8192} & $\sqrt{1/\ln N}$ & 0.2243 & 0.873 & 2.50$\times$$10^{-4}$ & 1.056 \\  
                               &  & & $\sqrt{4/\ln N}$   & 0.1717  & 2.536 & 2.47$\times$$10^{-4}$ & 1.207 \\  
                               &  & & $\sqrt{8/\ln N}$   & 0.1321  & 1.429 & 1.53$\times$$10^{-3}$ & 1.376 \\  \hline
\end{tabular}
\end{table}

\begin{figure}[b]
\epsfxsize=9cm
\centerline{\epsfbox{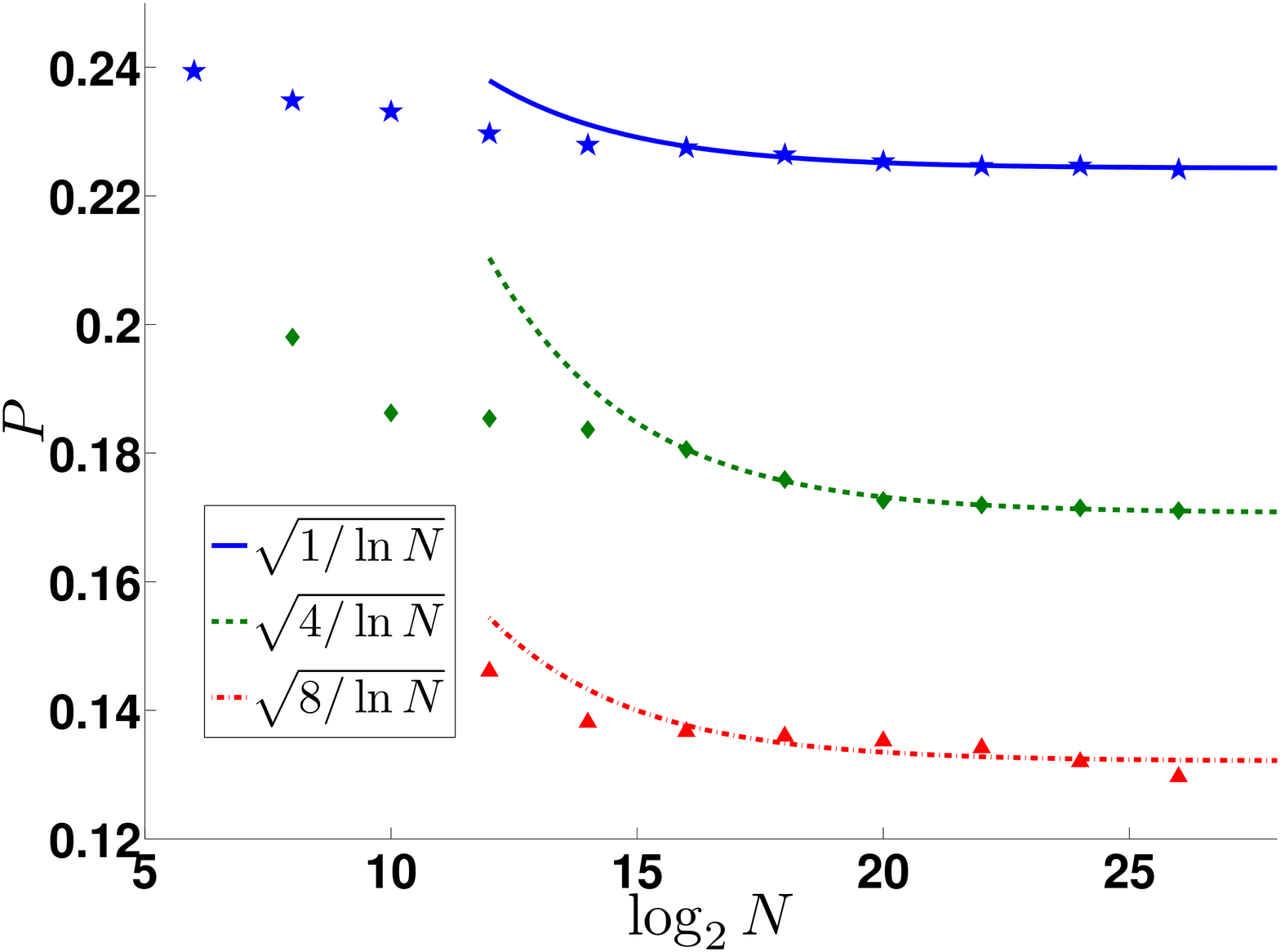}}
\caption{(Color online) Peak probability at the marked vertex as a function
of database size for different values of the ancilla control parameter
$\cos\delta$. We used $t_1=3$, and the curves are the fits
$P = a_1 + (b_1/L)$.}
\end{figure}

For a more elaborate comparison with Tulsi's analysis, we check whether
our fit parameters obey Eq.(\ref{PQvalues}) parametrised in terms of the
second moment $B$. Without ancilla control, values of $P$ and $t_2$ should
be related by $(4a_1)^{-1/2} = 8a_2/\pi$. That leads to the comparison
$0.31 \leftrightarrow 0.36$, which is quite reasonable. Stating it
another way, our estimates of $B$ vary from $0.31\sqrt{\log_2 N}$
to $0.36\sqrt{\log_2 N}$. Possible reasons for the discrepancy are:
(a) contribution of subleading terms neglected in Tulsi's analysis (e.g.
from the states outside the two-dimensional subspace used in the evolution),
(b) the notorious difficulty in accurately extracting parameters from
asymptotically logarithmic fits.

With ancilla control, the values of $B_\delta$ in Tables II and III match
well; control over infrared divergence definitely helps in extraction
of the scaling behaviour. They give estimates of $B$ varying from
$0.27\sqrt{\log_2 N}$ to $0.31\sqrt{\log_2 N}$. As per Eq.(\ref{optdelta}),
therefore, the optimal choice for the ancilla control parameter would be
\begin{equation}
(\cos\delta)_{\rm opt} \approx 3.5\sqrt{1/\log_2 N} \approx \sqrt{8.5/\ln N} ~.
\end{equation}
Our numerical simulations had database sizes varying from $N=2^{12}$ to
$2^{26}$. So in order to maintain $\cos\delta\leq1$, we had to restrict
the proportionality constant between $\cos\delta$ and $\sqrt{1/\ln N}$.
Nevertheless, our largest parameter choice, $\cos\delta=\sqrt{8/\ln N}$,
is close to optimal.

Finally, we combine the results for $P$ and $t_2$, to look at the scaling
behaviour of the algorithmic complexity $t_2/\sqrt{P}$. Without ancilla
control, the effective number of oracle calls scales as $\sqrt{N}\ln N$,
with $a_2/\sqrt{a_1}$ being the proportionality constant. Our results with
ancilla control are displayed in Fig.5, together with the fit parameters in
Table IV. The scaling of the effective number of oracle calls is improved
to $\sqrt{N\ln N}$, with the proportionality constant $a_3$ essentially
the same as $a_2/\sqrt{a_1}$. Due to the oscillatory pattern in the data
and non-negligible subleading corrections, our results are inadequate to
numerically optimise $a_3$. On the other hand, our estimates of $B$ and
Eq.(\ref{optcomplexity}) give $(a_3)_{\rm opt}\approx0.45$. That is
consistent with the value for our close to optimal parameter choice
$\cos\delta=\sqrt{8/\ln N}$, and hence we infer our best algorithmic
complexity to be $t_2/\sqrt{P} \approx 0.45 \sqrt{N\log_2 N}$.

\begin{table}[t]
\caption{Fit parameters for search oracle calls for various values of the
ancilla control parameter $\cos\delta$}
\vspace{2mm}
\begin{tabular}{|c|c|c|c|c|c|c|c|}\hline
 $s$ & $t_1$ & $L$ & $\cos \delta$ & $a_2$ & $b_2$ & Error & $B_\delta$ \\ \hline
\multirow{4} {*}{$1\over\sqrt{2}$} & \multirow{4} {*}3 & 512---16384 & 1 & 0.1412 & 2.755 & 2.25$\times$$10^{-3}$ & --- \\
 &  & 512---8192 & $\sqrt{1/\ln N}$ & 0.3463 & -2.782 & 2.41$\times$$10^{-3}$ & 1.059 \\
 &  & 512---8192 & $\sqrt{4/\ln N}$ & 0.2030 & -8.977 & 1.54$\times$$10^{-3}$ & 1.242 \\
 &  & 512---8192 & $\sqrt{8/\ln N}$ & 0.1562 &  3.290 & 3.81$\times$$10^{-3}$ & 1.351 \\  \hline
\end{tabular}
\end{table}

\begin{figure}[b]
\epsfxsize=9cm
\centerline{\epsfbox{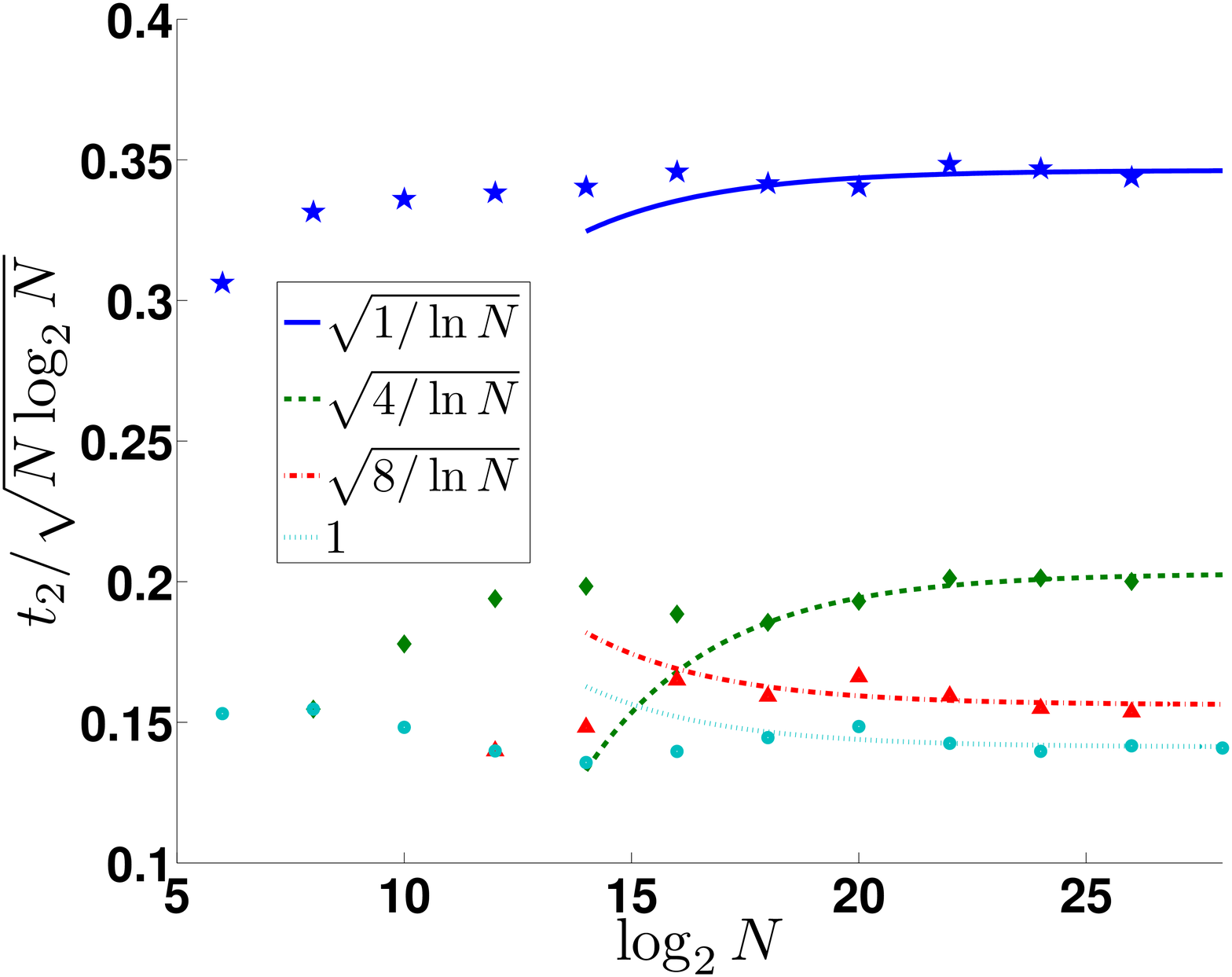}}
\caption{(Color online) Number of oracle calls as a function of database
size for different values of $\cos\delta$. We used $t_1=3$ and the curves
are the fits $t_2/\sqrt{N\log_2 N} = a_2 + (b_2/L)$.}
\end{figure}

\begin{table}[t]
\caption{Fit parameters for the algorithmic complexity for various values
of the ancilla control parameter $\cos\delta$}
\vspace{2mm}
\begin{tabular}{|c|c|c|c|c|c|c|}\hline
 $s$ & $t_1$ & $L$ & $\cos \delta$ & $a_3$ & $b_3$ & Error \\ \hline
\multirow{3} {*}{$1\over\sqrt{2}$} & \multirow{3} {*}3 & \multirow{3} {*}{1024---8192} & $\sqrt{1/\ln N}$ & 0.7336  & -13.06 & 5.31$\times$$10^{-3}$\\  
 &  &  & $\sqrt{4/\ln N}$ & 0.4911  & -24.30 & 4.03$\times$$10^{-3}$ \\  
 &  &  & $\sqrt{8/\ln N}$ & 0.4207  &  31.24 & 1.53$\times$$10^{-3}$ \\  \hline
\end{tabular}
\end{table}

\begin{figure}[b]
\epsfxsize=9cm
\centerline{\epsfbox{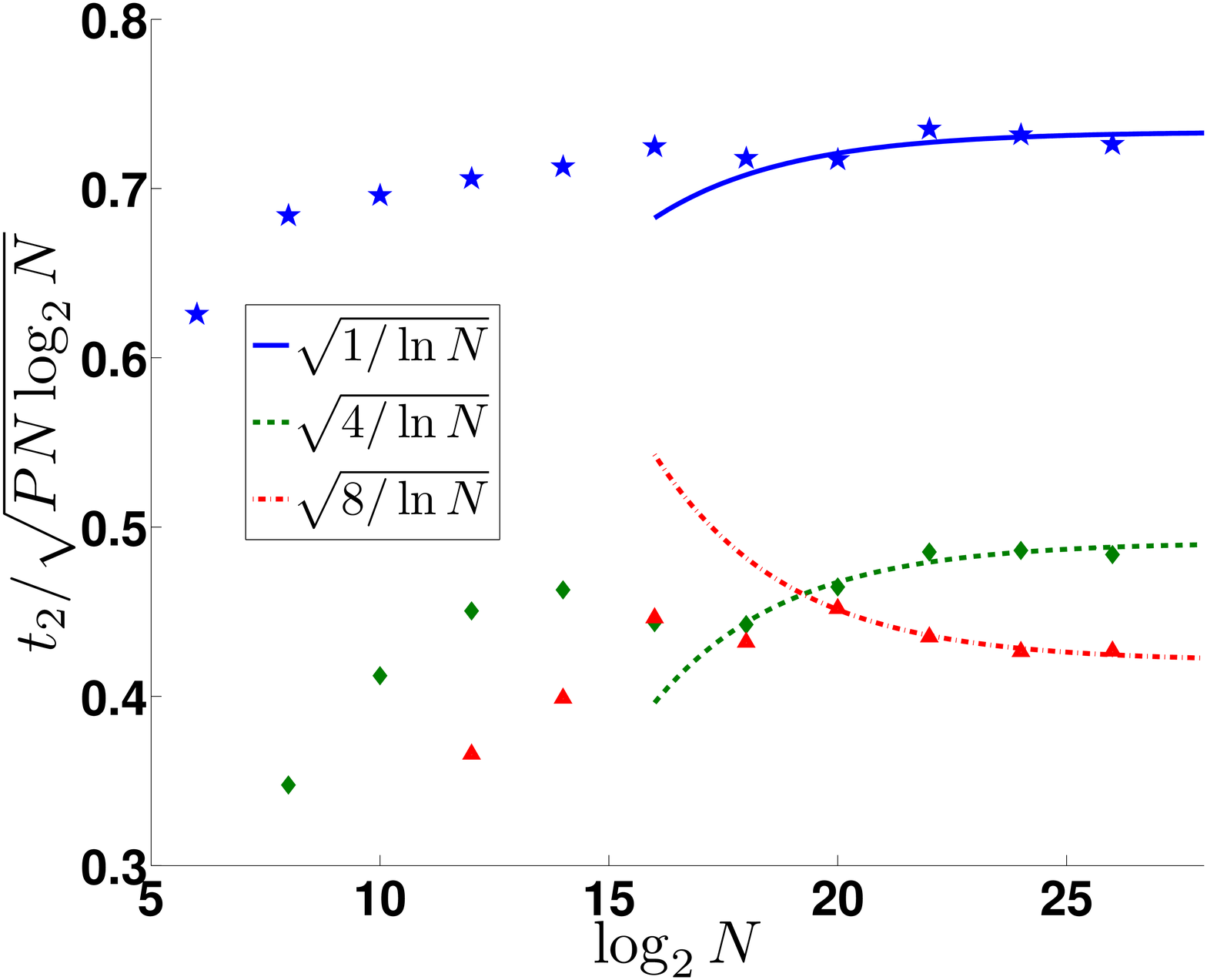}}
\caption{(Color online) Effective number of oracle calls as a function
of database size for different values of the ancilla control parameter
$\cos\delta$. We used $t_1=3$ and the curves are the fits
$t_2/\sqrt{PN\log_2 N} = a_3 + (b_3/L)$.}
\end{figure}

\section{Summary}

For the spatial search problem, $d=2$ is the critical dimension where
infrared divergences appear. Introduction of a mass term in the evolution
operator can suppress infrared divergences, and we have reinterpreted
Tulsi's ancilla control of the spatial search algorithm as introduction
of an effective mass at the marked vertex.

Our numerical results demonstrate how ancilla control improves the
scaling behaviour of the spatial search algorithm in $d=2$. In particular,
they agree with Tulsi's predictions \cite{tulsi}, and validate his analysis
criterion that the evolution of the quantum state is largely confined to a
two-dimensional subspace of the $N$-dimensional Hilbert space. The change
in scaling of $P$ with ancilla control is a clear signal for suppression
of the infrared divergence. Asymptotic behaviour of $t_2$ does not change,
however. It retains the $\sqrt{\ln N}$ factor beyond the lower bounds in
Eq.(\ref{bounds}), indicating that some effect of the critical behaviour
survives. Generically, logarithmic factors cannot be fully eliminated in
critical dimensions for interacting models of statistical mechanics.
Still it is an open question, whether the logarithmic factors found in
Tulsi's algorithm correspond to the minimal extra cost to be paid for
the $d=2$ spatial search problem, or whether they can be reduced further.

\break


\begin{thebibliography}{99}
\bibitem{oscillatory}{We do not properly understand this, but it perhaps
         indicates a second relevant length scale in the problem.}
\end{thebibliography}
\end{document}